# From Custom Logic to APIs: Understanding and Recommending API Replacement Refactorings

BRIDGET NYIRONGO, School of Computer Science and Technology, Beijing Institute of Technology, China
YANJIE JIANG*, School of Computer Software, Tianjin University, China
YUXIA ZHANG, School of Computer Science and Technology, Beijing Institute of Technology, China
HUI LIU, School of Computer Science and Technology, Beijing Institute of Technology, China

Software refactoring is essential for maintaining code quality. However, API replacement refactoring, which replaces custom logic with API calls, remains underexplored. Existing refactoring tools provide limited support for detecting such opportunities because they rely on predefined templates and have difficulty capturing complex, multi-statement semantic equivalents. To address this limitation, we conduct the first empirical study of API replacement refactorings by mining 166,299 commits across six open-source Java projects and manually analyzing a curated subset of 1,800 commits, from which we identify 366 validated instances to characterize their scope, categories, and recurring patterns. Based on these insights, we propose AKIRA (Adaptive Knowledge Discovery and Retrieval), a hybrid framework that integrates pattern-deterministic heuristics with a refactoring-aware knowledge base to assess the practical feasibility of recommending API replacement refactorings. Our evaluation shows that AKIRA achieves 90% recall and 88% precision on a manually curated dataset. Furthermore, on the external RETIWA dataset, AKIRA significantly improves the state of the art by increasing recall from 21% to 81% and precision from 40% to 78%. These results demonstrate the effectiveness of combining static pattern matching with semantic reasoning to support the automation of recommending complex API replacement refactorings.

CCS Concepts: • **Software and its engineering → Software creation and management**; • **Software post-development issues**; • **Software quality assurance**;

Additional Key Words and Phrases: Software Refactoring, API Replacement, Recommendation

## 1 Introduction

Software refactoring is the practice of restructuring internal code without altering external behavior, which is essential for maintaining long-term software quality and consistency [18]. While many refactoring types have been extensively studied and catalogued, certain patterns remain underexplored in practice. One such pattern is API replacement, where verbose, handwritten logic is substituted with semantically equivalent calls to standard or third-party libraries [36]. Developers often implement bespoke logic that replicates existing functionality, such as manual `null` checks instead of `Objects` utilities, explicit collection iteration instead of using `Streams`, or verbose assertions like `assertTrue(!flag)` instead of `assertFalse(flag)`. Such custom implementations increase maintenance effort, reduce readability, and propagate technical debt. Replacing them with well-tested library APIs consolidates logic, reduces duplication, and aligns code with modern development practices [36, 50].

*Corresponding author

Authors' Contact Information: Bridget Nyirongo, School of Computer Science and Technology, Beijing Institute of Technology, Beijing, China, larjean89@outlook.com; Yanjie Jiang, School of Computer Software, Tianjin University, Tianjin, China, yanjiejiang@tju.edu.cn; Yuxia Zhang, School of Computer Science and Technology, Beijing Institute of Technology, Beijing, China, yuxiazh@bit.edu.cn; Hui Liu, School of Computer Science and Technology, Beijing Institute of Technology, Beijing, China, liuhui08@bit.edu.cn.

Despite the practical relevance of API replacement refactorings, their prevalence, the characteristics of the logic they target, and the feasibility of automated recommendation remain underexplored. Existing tool support in mainstream IDEs [7, 17] and automated refactoring approaches is limited. Most techniques focus on structural refactorings or API migration scenarios, providing limited support for cases where multi-statement, semantically equivalent logic can be replaced with concise API calls. Template-based methods, such as RETIWA [50], partially address this challenge by matching handwritten logic to curated patterns, but they remain constrained to method-level templates and may not generalize to context-dependent replacements. Other approaches adopt a library-provider perspective, detecting API evolution to support client migration (e.g., RefacLib [48]) or target high-level architectural patterns rather than implementation-level logic [47]. These limitations highlight the need for an empirical study to understand the occurrence and nature of API replacement refactorings in real-world projects, and to guide the design of effective recommendation techniques.

To address these limitations, we conduct the first empirical study of API replacement refactorings, examining their prevalence, characteristics, and automation feasibility. We mined 166,299 commits from six diverse open-source Java projects and manually analyzed a curated subset of 1,800 commits, identifying 366 validated API replacement refactoring instances used to characterize recurring structural and semantic patterns in practice. Our study shows that API replacement refactorings typically transform guarded, multi-step logic into standardized API calls, often involving assertion and null-safety utilities, while framework-specific APIs exhibit higher syntactic variability. Our analysis also reveals inherent limitations of purely rule-based approaches. In practice, API replacement refactorings frequently involve (i) high syntactic variability across framework-specific or infrequently used APIs, (ii) context-sensitive logic spanning multiple statements, and (iii) semantic equivalence, where multi-step implementations are consolidated into a single API call. These patterns cannot be reliably identified using static heuristics alone, and the evolving nature of APIs, where new libraries and idiomatic usages emerge without predefined mappings [1], poses a significant challenge to reliable code analysis and test maintenance [15].

Motivated by these empirical insights, we propose AKIRA (Adaptive Knowledge Discovery and Retrieval), a hybrid recommendation framework that combines rule-based detection with adaptive knowledge-driven reasoning. Unlike prior LLM-based approaches [4, 40] for refactoring that rely on fixed prompts or static rule templates, AKIRA constructs a refactoring-aware, multi-view knowledge base from project history and employs agent-guided retrieval with adaptive thresholding ($\tau$) to dynamically adjust sensitivity based on code context (e.g., test code versus production logic). By retrieving and interpreting project-specific refactoring knowledge, the framework enables the discovery of emerging and evolving API replacement patterns without requiring predefined library mappings. To assess both within-project and cross-project generalization, we evaluate AKIRA in a two-stage setting: first on held-out datasets from two open-source Java projects, where it achieves up to 90% recall and 88% precision, and then on the external RETIWA [50] dataset, where recall improves from a baseline of 21% to 81% and precision from 40% to 78%. These results demonstrate substantial generalization across datasets constructed under different assumptions and show that AKIRA can reliably identify custom logic that can be replaced with well-tested, standardized APIs, directly supporting software reliability, testing, and automated maintenance. In summary, the paper makes the following contributions:

- An empirical study that quantifies API replacement refactorings in real-world open-source Java projects, characterizes their scope and categories, and identifies structural and semantic features that can guide automated detection.

```
1   public void printElementDecl(...) {
2     assertTrue(headCalled);
3   - assertTrue(!tailCalled);
4   + assertFalse(tailCalled);
5   }
6
```

Listing 1. Expression-level in Ant

```
1   private void log(Throwable t) {
2   - StringWriter sw = new StringWriter();
3   - PrintWriter w = new PrintWriter(sw);
4   - t.printStackTrace(w); w.close();
5   - log(sw.toString(), MSG_ERR);
6   + log(StringUtils.getStackTrace(t), MSG_ERR);
7   }
8
```

Listing 2. Statement-level in Ant

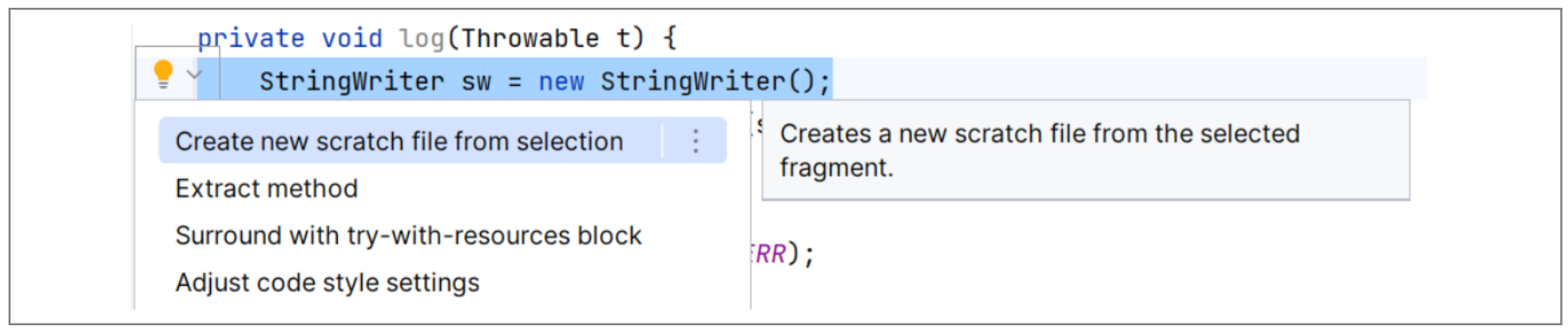

Fig. 1. IntelliJ IDEA quick-fix suggestions for Listing 2.

- A hybrid recommendation approach that integrates deterministic heuristics with an adaptive knowledge discovery and retrieval framework, enabling the identification of API-based replacement refactoring opportunities beyond common syntactic patterns.

The rest of the paper is structured as follows. Section 2 presents motivating examples for our study. Section 3 describes the three-phase methodology employed. Section 4 analyzes the frequency of API replacement refactorings. Section 5 examines the feature characteristics of API replacement refactorings. Section 6 assesses the practical feasibility of recommending API replacement refactorings by proposing a hybrid approach. Section 7 discusses threats to validity, core observations, and study implications. Section 8 reviews related work. Finally, Section 9 concludes the paper and outlines directions for future work.

## 2 Motivation Examples

In this section, we present two motivating examples to illustrate the distinct challenges involved in identifying opportunities for API replacement refactorings. The examples span a spectrum of complexity, ranging from simple syntactic substitutions to composite, multi-statement semantic replacements. Together, they highlight why a single analysis technique is insufficient and motivate the need for a hybrid detection and recommendation approach.

In **Listing 1**, the original code uses a negated boolean expression (`!tailCalled`) within an assertion, which obscures intent and complicates reading. The refactored version replaces it with `assertFalse`, a standard API that directly conveys behavior. This regular, local syntactic pattern can be effectively detected using deterministic heuristics and AST-based analysis

In **Listing 2**, the original code manually collects a `Throwable`'s stack trace using multiple low-level operations, reimplementing functionality already provided by libraries. The refactored version replaces this block with a single call to `StringUtils.getStackTrace(t)`. Unlike the local expression-level replacement in Listing 1, this composite, multi-statement refactoring spans disparate classes (`StringWriter`, `PrintWriter`) and requires semantic reasoning to recognize that it extracts a stack trace. This example illustrates the semantic gap that purely structural tools cannot bridge, motivating our hybrid detection approach.

For Listing 2, existing tools [7, 17] provide limited support for identifying API replacement opportunities. For example, IDEs such as IntelliJ IDEA [7] suggest local, structural quick fixes such as method extraction or try-with-resources (Figure 1) but do not address the underlying functional duplication. Similarly, approaches like RETIWA [50], which detect API-replacement opportunities ($m \rightarrow API$), rely on template-based clone detection at the method level. As a result, RETIWA struggles with composite refactorings where logic is fragmented across multiple statements and disparate classes. Thus, a need to shift the focus from syntax to underlying functional intent to align ad-hoc implementations with high-level API abstractions.

Table 1. Representative Examples of API Replacement Refactorings

| Refactoring Type | Source (Custom Logic) | Target (API Replacement) | Reference |
|---|---|---|---|
| **Expression Predicate** | `if (strLen == 0) return str;` | `if (isEmpty(str)) return str;` | CommonsLang [27] |
| **Control-flow Pattern** | `try op(); fail(); catch ( Exception e)` | `assertThrows(Exception.class, () -> op());` | Guava [19] |
| **Construction / Utility** | `new ArrayList<Object>() add( obj);` | `List.of(obj)` | Struts [6] |

These examples (Listing 1 and 2) illustrate that API-based refactoring requires a dual-layered approach: one recognizing structural idioms via deterministic detection, and another mapping custom logic to semantically equivalent APIs. This motivates AKIRA, our hybrid strategy. We employ deterministic heuristics for replacements of common patterns, coupled with the adaptive knowledge discovery and retrieval framework to recommend API replacement opportunities that static rules and templates cannot capture.

## 3 Methodology

### 3.1 Definition and Scope

Throughout this paper, we define *API replacement refactoring* as a behavior-preserving transformation in which developer-written code ranging from compound expressions to custom utility methods is replaced by semantically equivalent library API calls that subsume the original logic with a higher-level abstraction. We exclude (i) updates that only replace deprecated calls without removing or simplifying custom logic, (ii) purely structural code reorganizations, and (iii) changes that alter observable behavior. Table 1 presents representative replacements at multiple granularities, including expression-level checks, multi-statement control logic, and custom construction or utility logic.

### 3.2 Research Questions

We conduct a study to investigate the following research questions:

- **RQ1:** What is the frequency of API replacement refactorings across real-world open-source Java projects?
- **RQ2:** What kind of developer-written expressions or customized methods are more likely to be replaced with API calls?
- **RQ3:** Is it practical to automatically identify refactoring opportunities for API replacement?

This study is structured around three research questions that collectively investigate the frequency, characteristics, and detectability of API replacement refactorings in real-world open-source Java projects. RQ1 establishes the empirical relevance of this refactoring practice by examining the frequency with which developers replace custom logic with API calls. RQ2 identifies the structural and semantic code features that characterize such replacements, providing the analytical foundation for designing effective detection strategies. Building on these insights, RQ3 investigates the practical feasibility of automatically identifying API replacement refactoring opportunities, assessing whether a hybrid automated approach can reliably identify such opportunities in practice.

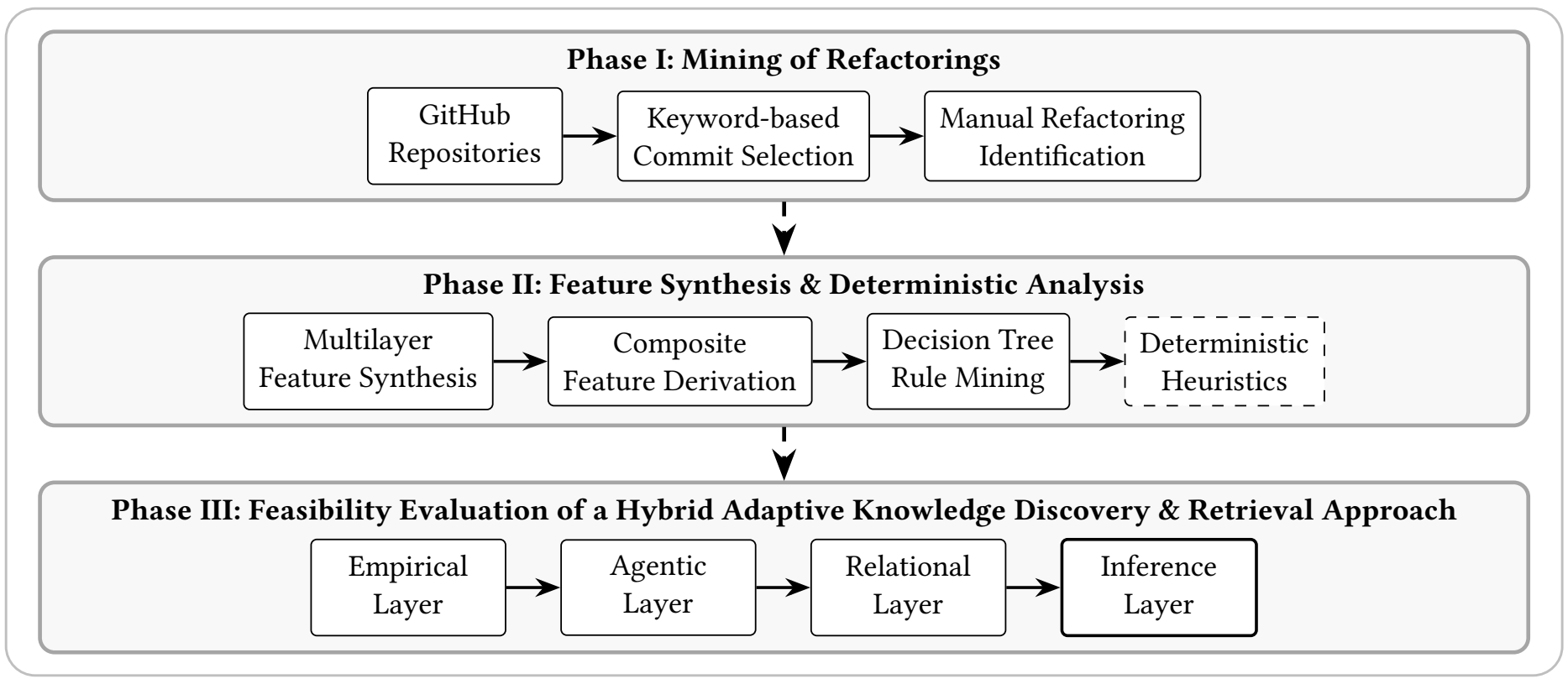

Fig. 2. Procedural Pipeline of the Three-Phase Study

### 3.3 Subject Applications

We collected and mined API replacement refactorings from six open-source Java projects following established guidelines for mining software repositories [26]. Candidate projects were identified on GitHub based on four criteria: (1) *Maturity,* active development within the last two years (2) *Community Impact,* at least 1,000 stars and 100 contributors, indicating substantial usage and peer scrutiny (3) *Complexity,* a minimum codebase size of 50 KLOC, and (4) *Domain Diversity,* coverage of varied application domains to reduce domain-specific bias. This process resulted in six widely used systems: *Guava* [19], *Spring Framework* [46], *Commons Lang* [27], *Struts* [6], *Elasticsearch* [16], and *Ant* [5]. Together, these projects span diverse architectures, from utility libraries to enterprise frameworks and distributed systems, providing a realistic context for studying API replacement refactorings in practice. Notably, we acknowledge that our findings may not fully generalize to proprietary codebases, which we leave for future investigation.

### 3.4 Procedural Pipeline

To investigate these questions, we employed a three-phase procedural pipeline, as illustrated in Figure 2. In *Phase 1*, we manually identified API replacement refactorings to answer RQ1, producing a ground-truth dataset of validated instances. *Phase 2* focused on feature synthesis and deterministic analysis for RQ2, where recurring structural and semantic features were extracted and composed into higher-level representations to induce a rule-based system capturing common, well-defined API replacement patterns. *Phase 3* evaluates the practical feasibility of automatically identifying API replacement refactorings by combining the deterministic heuristics from Phase 2 with an adaptive knowledge discovery and retrieval layer.

## 4 RQ1: Frequency of API Replacement Refactorings

### 4.1 Process

The process of mining API replacement refactorings from the subject applications consists of two main steps: keyword-based commit selection and manual identification of refactorings.

*4.1.1* ***Keyword-Based Commit Selection.*** We first identified candidate refactoring commits from the six open-source Java projects described in Section 3.3. Following a strategy adopted in prior studies [36, 44], we locally cloned each repository and filtered commits based on their commit messages using a predefined list of refactoring-related keywords drawn from existing work [2, 31, 44]. The complete list of keywords used in this study is available in our replication package [37].This keyword-based filtering step was necessary to reduce the manual inspection effort, as the six projects collectively contain approximately 166,299 commits. A commit was considered potentially relevant if its message contained at least one refactoring-related term, such as *refactoring*, *code cleanup*, *remove*, or *replace*. Applying this filter yielded 16,677 candidate commits for further analysis. As with prior keyword-based approaches, this process may not capture refactorings that are not explicitly mentioned in commit messages, such as floss refactorings [31]. Addressing this limitation is beyond the scope of this study. Future work may explore complementary techniques, including learning-based models, to improve recall.

*4.1.2* ***Manual Identification of Refactorings.*** We employed a two-stage manual process to identify API replacement refactorings as follows:

- *Commit Message–Based Filtering.* To prioritize commits likely to contain API replacements, we first identified candidate commits from the six open-source Java projects. From an initial pool of 16,677 keyword-matched commits, we constructed a statistically representative sample. A minimum sample size of 376 commits was required to achieve a 95% confidence level with a 5% margin of error. To ensure broader coverage across projects, we sampled 300 chronological matches per project ($n = 1,800$). This manual selection process was preferred over automated approaches to reduce misclassification and ensure the relevance of candidate commits. To assess potential selection bias introduced by keyword filtering over the full dataset (166,299 commits), we further validated the discarded set ($N = 149,622$). Using stratified random sampling, we selected 64 commits from the non-keyword-matched pool per project ($n = 384$). Manual inspection revealed only 3 API replacement instances, corresponding to a 0.78% false negative rate, indicating that the keyword-based heuristic captured the vast majority of relevant refactorings while keeping manual effort tractable.
- *Manual Code-Diff Inspection.* During this step two authors systematically analyzed the source code diffs of the 1,800 candidate commits. A commit was labeled as an API replacement refactoring only if it satisfied two criteria: (i) the removal or simplification of a handwritten expression or project-specific helper, and (ii) the introduction of a semantically equivalent standard library or third-party API, with no change to observable behavior. We excluded 1,649 commits that did not meet these criteria. Common false positives included *internal method extractions* (e.g., commit [41]), where logic was reorganized within the project rather than replaced by a higher-level API.

Notably, we also addressed *tangled commits*, i.e., commits containing multiple independent refactorings. For example, a single Guava commit [14] replaced multiple `try-catch-fail` blocks with the `assertThrows` API across several test classes. In such cases, each replacement was treated as a distinct refactoring instance to accurately capture the scope of API adoption. A change was considered an API replacement refactoring only if all unit and integration tests produced identical outputs before and after the modification, ensuring behavior preservation. To validate and ensure reliability, all candidate instances underwent dual-rater assessment: two authors, each with over eight years of professional software development and refactoring experience, independently evaluated semantic equivalence. The raters achieved a raw agreement of 95%, corresponding to approximately 60 disagreements. Each disagreement was reviewed by a third author who independently inspected the source code diffs, followed by joint discussions to reach consensus. This process yielded a Cohen's Kappa of 0.87, indicating substantial inter-rater agreement. Overall, the process resulted in 366 distinct API replacement instances extracted from 151 unique commits (see Table 2).

Table 2. Commit Retrieval and Discovery

| Project | #Proj. | #Ret. | #Anal | #FP | #TP | #Ref. |
|---|---|---|---|---|---|---|
| Guava | 7,134 | 5,459 | 300 | 280 | 20 | 90 |
| Spring | 34,231 | 2,594 | 300 | 283 | 17 | 33 |
| Commons | 9,068 | 811 | 300 | 267 | 33 | 91 |
| Struts | 8,142 | 460 | 300 | 270 | 30 | 41 |
| Elastic | 92,670 | 6,810 | 300 | 277 | 23 | 51 |
| Ant | 15,054 | 543 | 300 | 272 | 28 | 60 |
| **Total** | *166,299* | *16,677* | *1,800* | *1,649* | *151* | *366* |

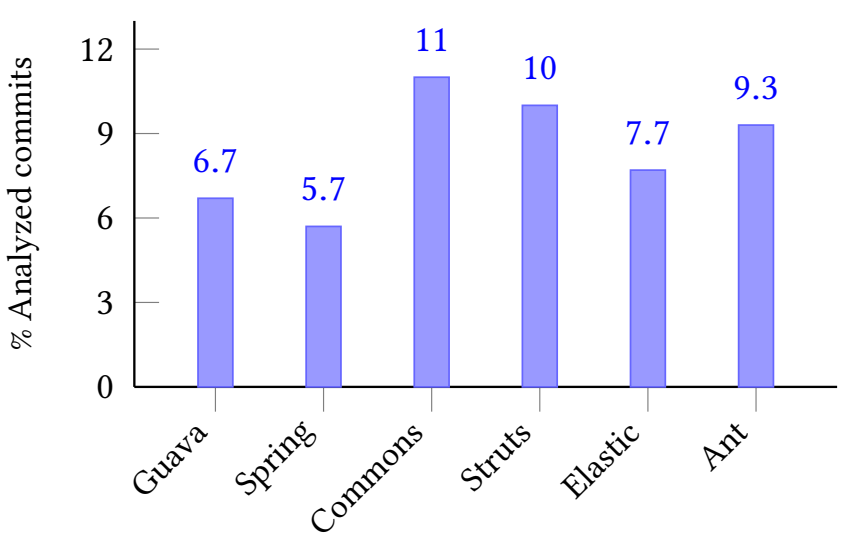

Fig. 3. API replacements (%).

## 4.2 Results and Analysis

*4.2.1 Frequency and Project Distribution.* From 16,677 commits retrieved using refactoring-related keywords, 1,800 were manually inspected, yielding 366 confirmed API replacement refactorings (Table 2). Overall, API replacement occurs in 8.4% of inspected commits. Per-project frequencies vary moderately: Guava (6.7%), SpringFramework (5.7%), CommonsLang (11%), Struts (10%), Elasticsearch (7.7%), and Ant (9.3%) (Figure 3). API replacements are concentrated in projects with heavy use of utility libraries (e.g, CommonsLang) rather than large framework-driven projects (e.g., SpringFramework). This pattern likely reflects the prevalence of repetitive, manual, or defensive code that benefits from declarative APIs (see Listing 3). Frequency is not determined solely by project size or age, and API replacements occur in both mature and actively evolving projects, though they are more common where custom logic is repetitive, fragile, or critical to core functionality.

```
1  public static String random(int count, String chars) {
2    /* Removed: if(chars==null)... return random(count, chars.toCharArray()); */
3    return secure().next(count, chars);
4  }
```

Listing 3. Commons Lang API replacement using declarative random string generation

*4.2.2 The Detectability Challenge of API Replacement Refactorings.* Keyword-based commit retrieval introduces substantial noise in identifying API replacement refactorings. From an initial set of 16,677 keyword-matched commits, we manually analyzed a sampled subset of 1,800 commits. Only 8.4% (151/1,800) contained at least one API replacement refactoring, yielding 366 validated instances and a 91.6% false-positive rate. This estimate is derived from a manually validated subset and reflects the prevalence of API replacement refactorings within the broader refactoring commit pool, where keyword filtering serves only as a recall-oriented preselection step. This high noise ratio indicates that most commits referencing APIs do not involve explicit logic replacement. Instead, API mentions often arise from usage expansion, API introduction, or unrelated maintenance changes. As a result, keyword-based matching alone is insufficient for reliably identifying API replacement refactorings, as these transformations are semantic in nature and often embedded in multi-step or context-dependent edits. Overall, API replacement refactorings remain difficult to detect using syntactic or keyword-based heuristics, highlighting the need for semantics-aware analysis.

**Algorithm 1:** Feature Extraction

```
   Input: Pos.snippets D_pos, Neg.snippets D_neg
   Output: Set of deterministic heuristics H
1  Initialize empty dataset D;
2  foreach snippet s with label y in D_pos ∪ D_neg do
3     F_struct ← Extract structural features via regex;
4     AST ← Parse s into an AST using javalang;
5     F_sem ← Extract semantic features from AST;
6     F_type ← Infer type-context from declarations;
7     Add ({F_struct, F_sem, F_type}, y) to D;
8  Train a decision tree classifier DT on D;
9  Initialize empty heuristic set H;
10 foreach decision path p in DT do
11    if confidence(p) ≥ 0.7 and support(p) ≥ 5 then
12       h ← Translate p into human-readable rule;
13       Add h to H;
14 return H
```

$\Longrightarrow$

Table 3. Top 8 Induced Heuristic Rules

| ID | Induced Heuristic Rule |
|---|---|
| H1 | $has_if \wedge sem_comp > 2 \wedge has_null_check$ |
| H2 | $has_new \wedge assertion_calls_present$ |
| H3 | $has_if \wedge sem_comp \leq 2 \wedge has_str \wedge has_null$ |
| H4 | $has_return \wedge has_null_check \wedge has_try$ |
| H5 | $has_new \wedge has_null_check$ |
| H6 | $ctrl_comp \geq 1 \wedge has_try$ |
| H7 | $has_return \wedge has_null_check \wedge has_new$ |
| H8 | $ctrl_comp \geq 1 \wedge has_try \wedge has_null_check$ |

**Finding 1**: API replacement refactorings are a substantial practice in Java software evolution, occurring in 8.4% of sampled commits. These refactorings span both popular and project-specific APIs. The high false-positive rate (91.6%) from keyword-based retrieval confirms that such replacements are semantics-driven and often hidden within complex commit histories.

## 5 RQ2: Features of API Replacement Refactorings

### 5.1 Process

To identify features that signal API replacement refactorings, we constructed a balanced dataset of 732 code snippets, consisting of the 366 manually validated API replacement instances from Section 4 and 366 negative instances sampled from the same subject applications that did not undergo replacement. We developed an automated feature-extraction pipeline (Algorithm 1) that derives 36 features across three dimensions:

- **Structural features**, capturing control-flow constructs such as null checks, early returns, and conditional branching.
- **Semantic features**, derived via AST analysis to capture operation patterns, multi-step expressions, and composite logic.
- **Type-context features**, heuristically inferred to highlight common data types (e.g., `String`, `List`) involved in API replacements.

A decision tree classifier (maximum depth = 12) was trained on these features to model their interactions. High-confidence decision paths were translated into interpretable *IF–THEN* heuristics (Table 3), which serve as rule signals for identifying candidate API replacement refactorings in recurrent, pattern-driven cases.

**Comparison with Alternative ML Classifiers.** To evaluate the robustness of our decision-tree rules, we compared their performance with standard classifiers using the same features. As shown in Table 4, the decision tree achieved the highest overall accuracy (0.787) and precision (0.974), though with moderate recall (0.587). Random Forests offered slightly higher recall (0.683) but lower precision (0.768) and overall accuracy (0.740), while SVM performed poorly across all metrics. These results demonstrate that our decision-tree rules provide a precise and interpretable method for identifying predictive features in the dataset.

Table 4. Performance comparison of ML classifiers on feature dataset.

| Classifier | Accuracy | Precision | Recall | F1 |
|---|---|---|---|---|
| Decision Tree | 0.787 | 0.974 | 0.587 | 0.733 |
| Random Forest | 0.740 | 0.768 | 0.683 | 0.723 |
| SVM | 0.496 | 0.491 | 0.444 | 0.467 |

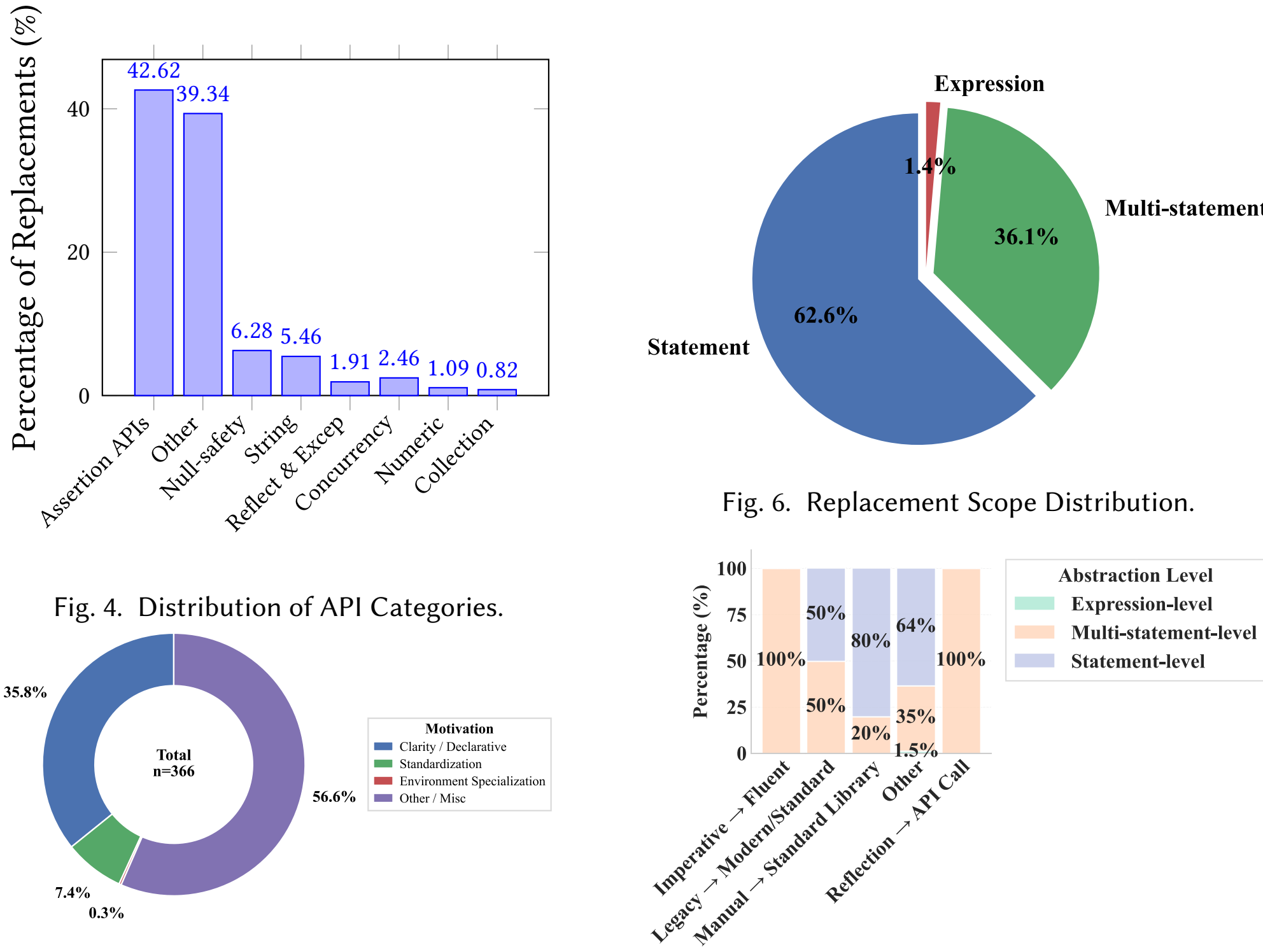

Fig. 4. Distribution of API Categories.

Fig. 5. Motivations for Replacement

Fig. 6. Replacement Scope Distribution.

Fig. 7. Abstraction levels across transition types.

## 5.2 Results and Analysis

To ensure the reliability of our qualitative analysis, the categorization of motivations, abstraction levels, and replacement scopes followed a rigorous manual annotation process. Two authors independently labeled each API replacement instance using a predefined coding scheme. Disagreements were resolved through discussion with a third author. The inter-rater agreement, measured using Cohen's Kappa, was $\kappa = 0.87$, indicating high annotation reliability. Our analysis reveals that API replacement refactorings are characterized by a semantic shift from multi-statement imperative blocks to concise, expression-level API calls. This transition is driven by recurring functional domains and structural patterns that can serve as signals for automated detection.

*5.2.1 Functional Domains and Motivations.* Figure 4 shows that assertion APIs constitute the largest category of API replacement refactorings (42.62%), followed by framework-specific APIs (categorized as *other*, 39.34%), null-safety utilities (6.28%), and string utilities (5.46%). This distribution emerged naturally from the empirical data rather than from methodological bias, reflecting the frequent evolution of test-related and framework-dependent logic. Although assertion APIs represent the single largest category, the majority of replacements (57.38%) occur in non-assertion domains, indicating that API replacement refactorings are not limited to testing code but also extend to broader production-level concerns such as validation, utility handling, and framework integration.

Figure 5 illustrates the primary drivers for these replacements, with the most dominant motivation being the need to improve clarity and adopt declarative coding styles (56.6%). While a variety of miscellaneous factors account for 35.8% of the cases, standardization remains a key explicit goal, representing 7.4% of the replacements. In many cases, verbose imperative implementations are replaced with concise, intent-revealing API calls that better communicate developer intent and align with established library conventions. Listing 4 illustrates this pattern: a manual diagnostic reporting sequence—comprising direct output to the error stream and explicit stack-trace printing—is replaced by a single declarative logging API call. While the original implementation focuses on low-level reporting logic, the refactored version delegates this responsibility to a logging utility that encapsulates formatting and severity management. Similar transformations appear in validation logic and defensive checks, where APIs encode reusable domain-specific conventions. Beyond reducing code size, these replacements improve maintainability by centralizing functionality within well-tested library abstractions, reducing cognitive load, and promoting standardized implementations.

```
1 - System.err.println(msg);
2 - e.printStackTrace();
3 + log(StringUtils.getStackTrace(e), Project.MSG_WARN);
```

Listing 4. Custom error-handling logic replaced by declarative logging API, reducing 2 lines to 1 in Ant

*5.2.2 Structural and Semantic Patterns.* Analysis of abstraction levels (Figure 7) and replacement scope (Figure 6) shows that API replacement refactorings predominantly operate at the statement (62.6%) and multi-statement level (36.1%), with expression-level replacements being rare (1.4%). This distribution indicates that developers mostly replace cohesive blocks of imperative logic rather than isolated expressions. Listing 5 exemplifies this pattern: an eight-line try–finally block implementing manual state management around factory invocation is collapsed into a single declarative API call, `instantiateWithFactoryMethod`, which encapsulates the same semantic intent. Similar transformations are observed for guarded control flow and resource or state management logic, where multiple statements collectively implement a single conceptual operation. The dominance of statement- and block-level replacements suggests that API replacement refactorings are driven by semantic consolidation rather than syntactic simplification, emphasizing the need for refactoring techniques that reason over control flow and execution context rather than isolated expressions.

```
1 - Method priorInvokedFactoryMethod = SimpleInstantiationStrategy.getCurrentlyInvokedFactoryMethod();
2 - try {
3 -       SimpleInstantiationStrategy.setCurrentlyInvokedFactoryMethod(method);
4 -       return beanSupplier.get();
5 - }
6 - finally {
7 -       SimpleInstantiationStrategy.setCurrentlyInvokedFactoryMethod(priorInvokedFactoryMethod);
8 - }
9 + return SimpleInstantiationStrategy.instantiateWithFactoryMethod(method, beanSupplier::get);
```

Listing 5. Custom state logic replaced by declarative API, reducing 8 lines to 1 in Springframework

*5.2.3 Quantitative Signals and Heuristics.* The observed patterns in Subsubsections 5.2.1 and 5.2.2 are reinforced by our automated feature analysis (Algorithm 1), which shows that guarded control structures with non-trivial semantic complexity are the strongest quantitative signals of API replacement refactorings. Expression- and statement-level constructs combining conditional branching with multi-step logic are particularly predictive, reflecting developers' preference for replacing defensive or validation-heavy code with declarative APIs. Decision tree learning (algorithm 1) over these features yields deterministic heuristics that formalize these recurring scenarios. For example, heuristic *H1* ($has_if \land sem_comp > 2 \land has_null_check$) in Table 3 captures cases where manually guarded logic is consolidated into null-safety or validation APIs. While effective for frequent, structurally regular replacements, these heuristics struggle with the syntactic and semantic diversity of framework-specific APIs, highlighting the need for adaptive, semantics-aware analysis to complement rule-based detection.

Overall, API replacement refactorings systematically target multi-statement, defensive, and domain-specific logic to improve clarity, maintainability, and best-practice adherence. Concentration in assertion and framework-specific APIs shows common patterns where verbose or fragile code is consolidated into concise, declarative calls. While structural and semantic signals support high-confidence heuristics, framework heterogeneity requires adaptive, semantics-aware approaches, highlighting the importance of capturing both code structure and intent for effective automated support.

**Finding 2:** API replacement refactorings mainly target cohesive blocks of manual, defensive, or domain-specific logic, concentrated in assertion APIs (42.62%) and framework-specific APIs (39.34%). Most occur at the statement level (62.6%) or span multiple statements (36.1%) and are driven by semantic consolidation rather than syntactic simplification. They are characterized by guarded control structures, semantic complexity, and type context. From these signals, we derive 8 high-confidence heuristics, while semantic analysis is needed for heterogeneous framework-specific cases.

## 6 RQ3: Automatic Recommendation of API Replacement Opportunities

### 6.1 Approach

Automatically identifying API replacement refactorings is challenging because these replacements exhibit substantial structural and semantic diversity. Our empirical analysis (Section 4 and 5) shows that replacements span single-statement expressions to multi-statement blocks and involve both common utility APIs and framework-specific APIs. While deterministic heuristics capture frequent, syntactically stable cases (e.g., assertion or null-safety APIs), they often miss context-dependent, multi-statement, and semantically rich replacements. Conversely, standalone knowledge-driven or LLM-based approaches can handle complex equivalences but may sacrifice precision, producing unstable or hallucinated recommendations [29]. These observations motivate a hybrid design that separates heuristic detections from cases requiring adaptive semantic reasoning, enabling reliable and generalizable automated detection.

To address the observed dynamics, we introduce AKIRA, a hybrid framework for the automatic identification of API replacement refactorings (Figure 8). Instead of relying on a single strategy, AKIRA combines deterministic detection (Tier 1) for predictable cases with adaptive semantic reasoning for structurally diverse or non-trivial replacements (Tier 2). The following subsections describe each tier in detail.

*6.1.1 Tier 1: Deterministic Pattern-Based Detection.* Tier 1 is a layer that identifies frequent and structurally well-defined API replacement opportunities using eight deterministic heuristics derived from Section 5. Each heuristic is implemented as a localized predicate over the Java Abstract Syntax Tree (AST), capturing recurring structural signals such as control-flow shape, method invocation patterns, and literal usage. We operationalise these heuristics as IDE-level quick-fix rules in an Eclipse [17]-based plugin prototype, where each heuristic maps to a quick-fix suggestion (e.g., via `IQuickFixProcessor`), similar to built-in refactorings such as *Convert loop to stream*. For each analyzed code snippet, Tier 1 evaluates the relevant AST context against the heuristic set. When a match is found, the framework generates a concrete API replacement candidate with an associated transformation template, which is exposed as a manually applicable quick fix. These transformations are safely applied using Eclipse JDT's `ASTRewrite` mechanism, and the tool is available in our replication package [37].

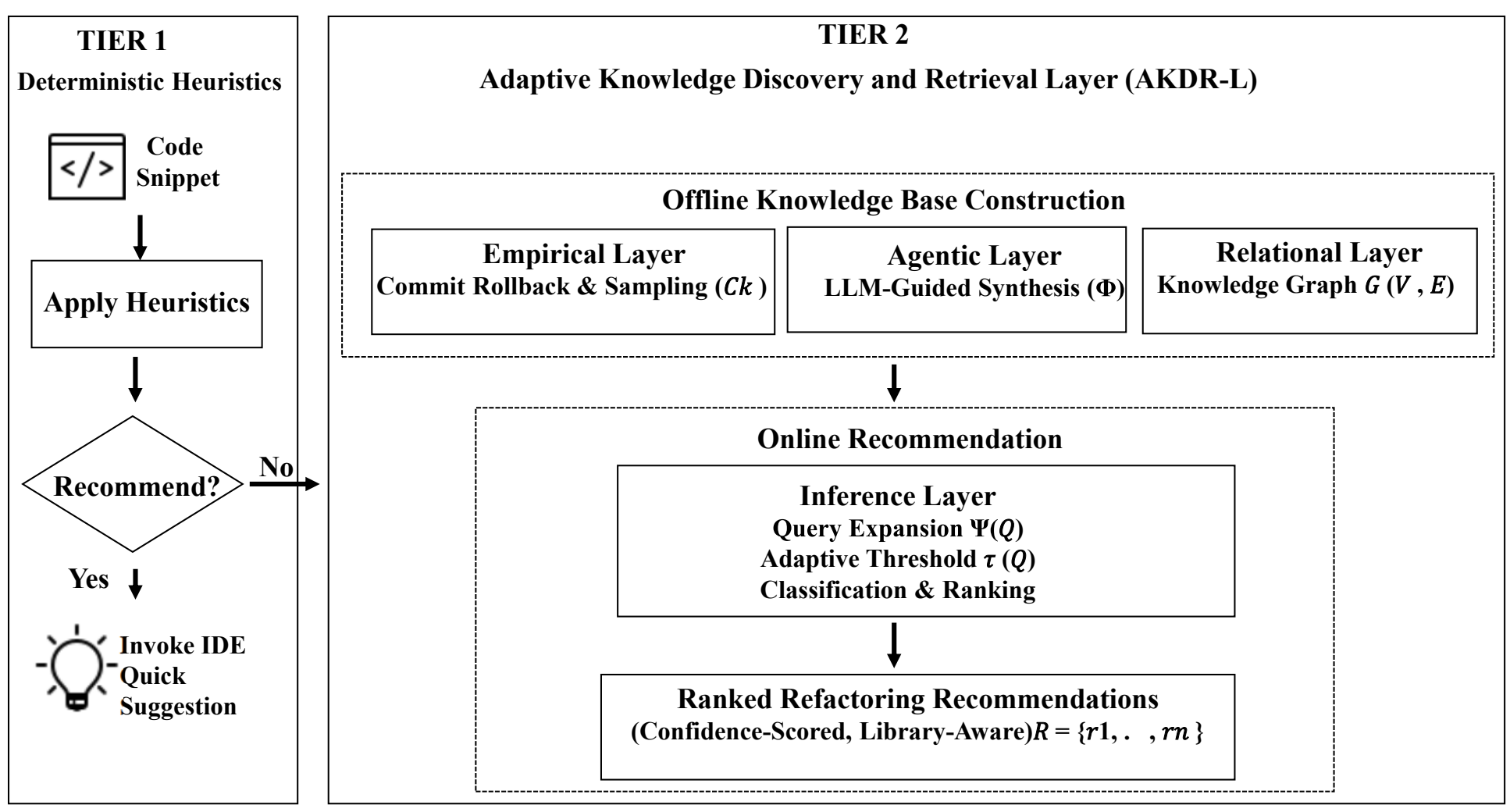

Fig. 8. AKIRA: a hybrid framework for API replacement using heuristics and AKDR-L.

*6.1.2 Tier 2: Adaptive Knowledge-Discovery and Retrieval Layer (AKDR-L).* AKDR-L handles cases not captured by Tier 1 by transforming raw Java code into concrete API replacement refactoring suggestions through four synergistic layers as presented in Algorithm 2. Each layer addresses a distinct challenge in identifying semantically equivalent replacements that elude deterministic heuristics.

*6.1.3 Empirical Layer: Multi-Library Temporal Sampling.* Given the challenge of analyzing large-scale multi-library repositories, AKDR-L employs a temporal rollback mechanism with adaptive sampling. For each library $L_k$ in $\mathcal{L} = \{L_1, L_2, \ldots, L_6\}$, we establish a cutoff date $t_c$ and retrieve the corresponding commit. That is, $C_k$ denotes the most recent commit of $L_k$ whose timestamp does not exceed the cutoff date $t_c$, as follows:

$$C_k = \arg \max_{c \in \text{Commits}(L_k),\, \text{Date}(c) \leq t_c} \text{Date}(c) \tag{1}$$

The sampling process follows library-aware proportional allocation:

$$S_k = \min\left(N_k, \left\lfloor \rho \cdot \frac{N_k}{\sum_{j=1}^{|\mathcal{L}|} N_j} \cdot M_{\text{total}} \right\rfloor\right) \tag{2}$$

where $N_k$ is the total number of methods in library $L_k$, $\rho = 0.3$ is the global sampling rate, and $M_{\text{total}} = 2000$ is the token budget (pilot-tested for semantic saturation), ensuring proportional library coverage while respecting the LLM's context window and computational limits.

*6.1.4 Agentic Layer: Multi-View Knowledge Synthesis (Φ).* Rather than treating code as raw text, AKDR-L employs Tree-Sitter [9] to extract Concrete Syntax Trees (CSTs). A specialized Knowledge

**Algorithm 2:** Adaptive Knowledge-Discovery and Retrieval Layer (AKDR-L)

```
Input: Multi-library repository set R = {R_1, R_2, ..., R_6}, User query Q
Output: Refactoring Recommendation context S_final with library attribution
1  Phase I: Offline Knowledge Base Construction
2  foreach library L_k ∈ L do
3      R'_k ← TemporalRollback(R_k, t_c) // Empirical layer
4      foreach method m_i ∈ Sample(R'_k, S_k) do
5          ω_i ← Parse(m_i) using TreeSitter;
6          K_i ← Φ(ω_i, L_k) // Agentic layer: multi-view synthesis
7          Update graph G and vector store V with K_i;
8  Phase II: Online Adaptive Retrieval
9  Q_exp, L_Q, T_Q ← Ψ(Q) // Library-aware expansion
10 τ ← SelectThreshold(T_Q, L_Q) // Adaptive thresholding
11 S_graph ← GraphSearch(G, Q_exp, τ, L_Q) // Relational layer
12 if max(score(S_graph)) < τ then
13     S_candidates ← VectorSearch(V, Embed(Q_exp), L_Q);
14 else
15     S_candidates ← S_graph;
16 Phase III: Pattern-Aware Synthesis with Library Context
17 S_reranked ← LLM_Library_Aware_ReRank(Q, S_candidates, L_Q);
18 S_final ← AdaptiveClassification(S_reranked, T_Q, L_Q);
19 return S_final
```

Architect agent applies a transformation function $\Phi$ to every extracted method $m_i$ to produce an enhanced structured knowledge tuple $K_i$:

$$K_i = \Phi(m_i, L_k) = \langle f_i, \mathcal{P}_i, \mathcal{T}_i, \mathcal{L}_i, \mathbf{v}_i \rangle \tag{3}$$

where:

- $f_i$ is the functional purpose inferred via structured reasoning over the parsed code
- $\mathcal{P}_i = p_1, p_2, \ldots$ is the set of detected refactoring patterns, building upon the feature taxonomy from RQ2
- $\mathcal{T}_i \subseteq$ test, utility, business is the code type classification
- $\mathcal{L}_i \in \mathcal{L}$ is the source library context, critical for library-specific pattern recognition
- $\mathbf{v}_i$ is the library-aware embedding vector: $\mathbf{v}_i = \text{Embed}(L_i \parallel f_i \parallel \mathcal{P}_i \parallel \mathcal{T}_i)$

The synthesis incorporates library-specific pattern dictionaries $\mathcal{D}_L$ for each library $L$, enabling targeted detection of framework-specific improvement opportunities.

*6.1.5 Relational Layer: Pattern-Centric Knowledge Graph (G).* We construct a multi-library knowledge graph $G = (V, E)$, where each node $v \in V$ represents a structured knowledge tuple $K_i$. Edges $E$ capture semantic relationships based on shared refactoring targets rather than textual or structural similarity. Specifically, two nodes are connected if the same API replacement strategy can be applied to both snippets, independent of their original library, syntax, or surface-level structure.

The edge weights are defined by a multifaceted similarity function that accounts for architectural, functional, and contextual relationships:

$$
\begin{aligned}
w(K_i, K_j) = \alpha \cdot \mathbb{I}(\mathcal{L}_i = \mathcal{L}_j) & \quad \text{(library cohesion)} \\
+ \beta \cdot \text{Jaccard}(\mathcal{P}_i, \mathcal{P}_j) & \quad \text{(pattern alignment)} \\
+ \gamma \cdot \mathbb{I}(\mathcal{T}_i = \text{test} \wedge \mathcal{T}_j = \text{test}) & \quad \text{(test code affinity)} \\
+ \delta \cdot \text{Overlap}(Keywords_i, Keywords_j) & \quad \text{(conceptual overlap)}
\end{aligned} \tag{4}
$$

We set $\alpha = 0.6$, $\beta = 0.5$, $\gamma = 0.4$, and $\delta = 0.15$ to prioritize library context and refactoring pattern alignment over superficial similarity. By connecting code snippets based on shared API replacement potential, this graph enables cross-library refactoring target discovery, supporting the detection of semantically equivalent refactoring opportunities across different frameworks.

*6.1.6 Inference Layer: Adaptive Retrieval and Classification.* The inference layer is composed of three components, i.e., *Library-Aware Query Expansion*, *Adaptive Threshold Selection*, and *Adaptive Classification*. For an incoming query $Q$, *Library-Aware Query Expansion* component performs library detection and conceptual expansion:

$$Q_{\text{exp}} = \Psi(Q) = \langle C, \mathcal{P}_Q, \mathcal{L}_Q, \mathcal{T}_Q \rangle \tag{5}$$

where $\Psi$ uses both keyword matching and LLM analysis to identify core concept $C$, potential refactoring patterns $\mathcal{P}_Q$, relevant libraries $\mathcal{L}_Q$, and code type $\mathcal{T}_Q$. Building on observations from Section 5, the *Adaptive Threshold Selection* component employs context-aware retrieval thresholds:

$$
\tau(Q) = \begin{cases} 0.35, & \text{if } \mathcal{T}_Q = \text{test} \quad \text{(higher recall for test code)} \\ 0.40, & \text{if } \mathcal{L}_Q \neq \emptyset \quad \text{(library/internal improvements)} \\ 0.60, & \text{otherwise (Standard API)} \quad \text{(higher precision)} \end{cases} \tag{6}
$$

This adaptive approach addresses the variability identified in Section 5, where test code patterns often involve subtle API replacements requiring lower similarity thresholds for detection.

$$
\begin{aligned}
S_{\text{graph}} &= \text{GraphSearch}(G, Q_{\text{exp}}, \tau(Q)) \\
S_{\text{vector}} &= \begin{cases} \text{VectorSearch}(\mathcal{V}, Q_{\text{exp}}), & \text{if } \max_{s \in S_{\text{graph}}} \text{score}(s) < \tau(Q) \\ S_{\text{graph}}, & \text{otherwise} \end{cases} \\
S_{\text{final}} &= \text{LLM_ReRank}(Q, S_{\text{vector}}, \mathcal{L}_Q)
\end{aligned} \tag{7}
$$

*Adaptive Classification* component employs an adaptive classification function $\Lambda$ that combines raw LLM output with contextual signals:

$$\text{Replace}(Q) = \Lambda(\text{LLM}_{\text{raw}}(Q), \mathcal{T}_Q, \mathcal{L}_Q, \text{Confidence}(Q)) \tag{8}$$

where $\Lambda$ applies type-specific confidence thresholds and library-aware pattern validation to reduce false negatives while maintaining precision.

*6.1.7 Parameter Sensitivity and Design Choices.* The design parameters in AKDR-L, including retrieval thresholds ($\tau$), edge weights ($\alpha, \beta, \gamma, \delta$), and token budget ($M_{\text{total}}$), were determined through systematic sensitivity analysis on a validation subset. Specifically, $\tau$ was explored over the range [0.20, 0.80] (step = 0.05), analyzing precision–recall trade-offs. The results show stable performance across this range, yielding a mean F1-score of 0.89 with low variance ($\sigma = 0.024$), and final thresholds were chosen from stable inflection regions. Similarly, perturbing edge weights by ±20% resulted in performance variations within ±3.0%, indicating robustness to weight configuration. For the token budget, we evaluated values in increments of 1,000 tokens and observed clear semantic saturation at $M_{\text{total}} = 2000$: increasing the budget from 1,000 to 2,000 improved accuracy by 7.4%, whereas further increasing to 3,000 yielded less than 1.0% improvement while increasing latency by 42%. These results justify the selected configuration as a balance between effectiveness and efficiency. Finally, the hybrid architecture itself is motivated by empirical observations (Section 5), where deterministic heuristics effectively capture simple patterns, while semantically complex replacements require adaptive retrieval.

*6.1.8 Implementation.* AKDR-L is implemented as a fully automated Python pipeline that recommends API replacement refactorings by indexing and reasoning over API replacement instances mined from the subject applications described in Sections 3.3. To ensure structural precision, the framework utilizes Tree-Sitter [9] for concrete syntax tree (CST) parsing, facilitating the extraction of code patterns even from partial snippets. The LLM-based Agentic Layer leverages GPT-4o [38] with deterministic decoding ($T = 0$) to synthesize refactoring metadata into structured JSON. GPT-4o was chosen over other models,e.g., LLaMA [49], to provide the advanced semantic reasoning necessary for cases involving multi-statement refactorings, which smaller, less generalized models cannot handle. To mitigate inherent LLM non-determinism, all outputs are passed through a post-processing validator that uses Tree-Sitter to verify syntactic and type correctness before presenting the recommendation as a verified hypothesis. The Retrieval and Relational Layers employ a hybrid indexing strategy: dense retrieval is handled via 1,536-dimensional code embeddings (text-embedding-3-small) indexed using FAISS [24], while relational reasoning is performed through a multi-library knowledge graph in NetworkX [20]. The inference loop integrates automated query expansion and adaptive thresholding to enable generalization across diverse contexts. The current implementation operates as an external, on-demand analysis pipeline with all components released as part of the replication package [37].

## 6.2 Evaluation

*6.2.1 Experimental Setup.* We evaluate our approach through a two-step process targeting internal validity and external generalization:

**Manually curated dataset:** We constructed a ground-truth dataset from two held-out Java projects, JUnit 5 [25] and Quarkus [43], which were excluded from feature mining, heuristic derivation, and knowledge-base construction. The dataset was created through a fully manual labeling process without relying on any automated detection tools or heuristics from our approach. We first applied keyword-based filtering to identify candidate commits potentially containing API replacement instances, yielding an initial pool of commits from both projects. To ensure temporal independence, only commits created after a knowledge-base cutoff date of December 31, 2024 were retained. The resulting candidates were then manually inspected at the code-diff level by two authors, who independently labeled each instance. Disagreements were resolved through discussion with a third author, achieving a Cohen’s Kappa score of 0.87, indicating strong agreement. This process yielded 58 validated API replacement instances, alongside 58 non-replacement instances sampled from the same pool to construct a balanced dataset of 116 instances. To prevent data leakage, we enforced triple isolation: (i) *Structural*, evaluating only projects absent from the knowledge base. (ii) *Temporal*, separating the knowledge-base construction period from the evaluation data. And (iii) *API Membership*, where 97% of API calls in the evaluation set are absent from the knowledge base. These measures ensure that AKIRA generalizes to previously unseen refactoring scenarios, closely approximating real-world deployment conditions.

**External Generalization (RETIWA Dataset):** We also evaluate on the RETIWA dataset [50], containing 337 API-replaceable code instances. To construct a balanced set, we supplement these with 337 non-candidate fragments, enabling assessment of AKIRA’s generalization to external, third-party codebases.

Table 5. Detection performance on the manually curated dataset (balanced, post-knowledge-base commits).

| Approach | TP | FN | TN | FP | Recall | Precision | Accuracy |
|---|---|---|---|---|---|---|---|
| **AKIRA** | **52** | **6** | **51** | **7** | **90%** | **88%** | **89%** |
| Deterministic Heuristics | 14 | 44 | 51 | 7 | 24% | 67% | 56% |
| ChatGPT-4o (Vanilla baseline) | 18 | 40 | 45 | 13 | 31% | 58% | 54% |
| AKDR-L | 50 | 8 | 33 | 25 | 86% | 67% | 72% |
| RETIWA-baseline | 2 | 56 | 54 | 4 | 3% | 33% | 5% |

Table 6. Detection performance on the external RETIWA dataset (balanced).

| Approach | TP | FN | TN | FP | Recall | Precision | Accuracy |
|---|---|---|---|---|---|---|---|
| **AKIRA** | **272** | **65** | **261** | **76** | **81%** | **78%** | **79%** |
| Deterministic Heuristics | 64 | 273 | 261 | 76 | 19% | 46% | 48% |
| ChatGPT-4o (Vanilla baseline) | 114 | 223 | 253 | 79 | 34% | 59% | 55% |
| AKDR-L | 269 | 68 | 93 | 244 | 80% | 52% | 54% |
| RETIWA-baseline | 70 | 267 | 232 | 105 | 21% | 40% | 45% |

Table 7. Ablation study: contribution of individual AKDR-L components.

| Configuration | Change/Removal | Acc. | Prec. | Rec. | F1 | Spec. |
|---|---|---|---|---|---|---|
| **Default (Full)** | *Complete AKDR-L* | **72.0%** | **67.0%** | **86.0%** | **75.3%** | **58.2%** |
| w/o LLM reranking | Removes contextual re-scoring | 69.0% | 62.7% | 94.0% | 75.2% | 44.0% |
| w/o library awareness | Removes project pattern dicts. | 68.0% | 62.2% | 92.0% | 74.2% | 44.0% |
| w/o query expansion | Disables conceptual enrichment | 67.0% | 62.0% | 88.0% | 72.7% | 46.0% |
| w/o knowledge graph | Replaces KG with vector sim. | 61.0% | 57.3% | 86.0% | 68.8% | 36.0% |
| w/o adaptive classif. | Uses fixed $\tau$ (no context) | 52.0% | 51.0% | 100.0% | 67.6% | 4.0% |

*Acc. = Accuracy, Prec. = Precision, Rec. = Recall, F1 = F1-score, Spec. = Specificity.*

*6.2.2 Results and Analysis.* We evaluate five configurations: (i) Deterministic Heuristics (Tier 1 baseline), (ii) AKDR-L (knowledge-driven Tier 2), (iii) AKIRA, which integrates AKDR-L with heuristic pre-filtering, (iv) ChatGPT-4o (vanilla baseline), and (v) a snippet-level adaptation of RETIWA [50]. Performance is measured using standard classification metrics (TP, FP, TN, FN), from which Precision, Recall, and Accuracy are computed. Results are organized to compare heuristic-only detection, knowledge-driven detection, general-purpose LLM reasoning, hybrid integration, and history-dependent baselines.

**Heuristic Baseline and Knowledge-Driven Detection**: Table 5 reports results on the manually curated dataset. Deterministic heuristics achieve relatively high precision (67%) but low recall (24%), confirming that purely syntactic rules capture a limited subset of API replacement opportunities. AKDR-L substantially improves recall to 86% by identifying semantically motivated replacements beyond fixed patterns. The ChatGPT-4o vanilla baseline provides an intermediate reference point, achieving 31% recall and 58% precision, indicating that general-purpose LLM reasoning can capture some API replacement opportunities but remains behind task-specific AKDR-L or hybrid AKIRA. The snippet-level RETIWA baseline attains only 3% recall, as it relies on historical change information unavailable in snippet-level settings, providing a lower-bound reference for history-dependent approaches.

**Hybrid Synergy Analysis**: AKIRA achieves the highest overall performance, reaching 90% recall and 88% precision on the manually curated dataset. By using deterministic heuristics as a pre-filter, AKIRA recovers the majority of semantically rich refactorings identified by AKDR-L while reducing irrelevant and low-quality recommendations. Compared to the ChatGPT-4o vanilla baseline, AKIRA not only achieves substantially higher recall and precision but also demonstrates the advantage of task-specific, knowledge-driven reasoning over general-purpose LLM suggestions in structured API replacement detection.

**External Generalization on the Retiwa Dataset**: Table 6 evaluates generalization on the external Retiwa dataset. Deterministic heuristics remain conservative (19% recall), while AKDR-L improves coverage substantially (80% recall) but exhibits limited precision (52%). ChatGPT-4o achieves moderate performance (34% recall and 59% precision), indicating that general-purpose LLM reasoning provides partial coverage but may not be sufficient for fully reliable API replacement detection. AKIRA outperforms all standalone approaches, achieving 81% recall and 78% precision, indicating that heuristic filtering combined with knowledge-driven reasoning remains highly effective even in third-party codebases unseen during knowledge-base construction. The snippet-level Retiwa baseline attains only 21% recall and 40% precision, highlighting the challenge of identifying API replacement opportunities without historical change information.

**Ablation Study**: Table 7 presents an ablation study assessing the contribution of individual AKDR-L components. Removing any architectural layer degrades performance, confirming that the effectiveness of AKDR-L arises from component synergy rather than any single mechanism. Adaptive classification is most critical, disabling it results in 100% recall but collapses specificity to 4%, generating excessive false positives. Removing the knowledge graph or query expansion substantially reduces precision and F1, underscoring the importance of semantic enrichment for non-trivial refactorings. In contrast, LLM reranking and library awareness primarily act as refinement layers, improving precision and specificity rather than recall.

**Finding 3:** Automatic identification of API replacement refactorings is practical when using a hybrid framework. AKIRA achieves the highest overall performance, with 90% recall and 88% precision, improving coverage by 66% over standalone heuristics and precision by 21–26% over AKDR-L. Ablation analysis shows that Adaptive Classification is critical for controlling false positives, reducing them by 50%. These results demonstrate that combining deterministic heuristics with knowledge-driven retrieval enables accurate, snippet-level API replacement detection.

## 7 Discussion

### 7.1 Threats to Validity

**External Validity:** Our study focuses on six open-source Java projects. While diverse in domain, the findings may not generalize to proprietary systems or other programming languages. The dataset of 1,800 manually validated commits represents a subset of the broader refactoring landscape. Specifically, reliance on keyword-based discovery may omit *floss refactorings* [31] that occur alongside functional changes without explicit commit documentation. We mitigated these threats by selecting mature, long-lived projects such as Springframework [46] and Guava [19] and by employing a rigorous manual review process involving three authors. Finally, while LLM latency (15–30 minutes) remains a constraint, our asynchronous architecture ensures that recommendations are delivered without blocking the developer's real-time workflow.

**Internal Validity:** Potential threats include human error during the manual code-level analysis and selection bias from keyword filtering. To mitigate these, we operationalized behavior preservation using existing test suites. A refactoring was considered valid only if all tests passed pre- and post-modification. Candidate commits were independently validated by two authors, with a third resolving disagreements (Cohen's Kappa = 0.87). To address selection bias, we verified a sampled subset of non-keyword-matched commits, confirming a low false-negative rate. Finally, we enforced triple isolation—structural, temporal, and API-membership—to prevent data leakage between our knowledge base and the evaluation set.

**Construct Validity:** AKIRA may occasionally produce false positives or miss replacements that require inter-procedural reasoning due to its reliance on snippet-level context. Library-aware embeddings and LLM reranking partially mitigate this, but the black-box nature of the underlying models means some recommendations may still require developer verification. Additionally, comparing RETIWA at the snippet level may not fully reflect its intended usage. However, RETIWA represents the state of the art in API replacement, and including it as a baseline still provides meaningful context for evaluation. Future work could integrate formal verification to ensure that all suggested API replacements are behavior-preserving.

### 7.2 Practical Usability, Adaptability, and Limitations

AKIRA offers a scalable and practical approach to API replacement refactoring in real-world development environments by combining deterministic detection with semantic reasoning. Tier 1 heuristics provide immediate, precision-oriented suggestions that can be surfaced as *quick fix suggestions*, enabling developers to apply common, well-structured transformations with minimal cognitive overhead. Complex or context-dependent cases are deferred to the AKDR-L tier, which performs library-aware semantic reasoning to identify multi-statement or semantically rich replacements beyond the reach of static rules. Our evaluation (Table 5) shows that this hybrid design achieves high recall and precision while remaining practically applicable. Beyond reproducing developer-performed replacements, AKIRA can surface *latent refactoring opportunities*, such as removing redundant literals via overloaded API constructors (see Listing 6), which enhance readability and maintainability, even when labeled as false positives in the curated evaluation ground truth. At the same time, frontier challenges remain for composite replacements that span project-specific accessors and library utilities (e.g., Listing 7), which require inter-procedural data-flow reasoning and graph-based representations beyond snippet-level matching. By separating IDE detection from semantic inference, the framework aligns with real-world developer workflows while remaining adaptable to other IDEs (e.g., IntelliJ IDEA [7], VS Code [30]) and programming paradigms through interchangeable parsing and embedding components.

```
1 - return new PathMatch<>(path, "", next.getValue());    // original constructor with empty string
2 + return new PathMatch<>(path, next.getValue());        // simplified via overloaded constructor
```

Listing 6. Quarkus snippet illustrating a latent API replacement

```
1 - boolean isImportPolicy() { return importPolicy; }       // removed method
2 - sentryTool.isImportPolicy()                             // removed usage
3 + !StringUtils.isEmpty(sentryTool.getImportPolicyFilePath()) // added composite expression
4 + StringUtils.isEmpty()                                    // added API method
5 + getImportPolicyFilePath()                                 // added custom getter
```

Listing 7. Undetected API replacement involving a composite expression from RETIWA dataset

### 7.3 Implications

**For Practitioners:** Automating the transition from custom logic to standardized library APIs is vital for improving software reliability. Integrating AKIRA-like agents into IDEs [17] or CI/CD pipelines [32] enables proactive modernization by subsuming error-prone logic into tested APIs. This reduces maintenance overhead and acts as a pedagogical bridge, helping developers adopt idiomatic patterns and enforce standards via automated logic simplification.

**For Researchers:** This work shifts refactoring mining toward semantic logic subsumption, aligning with neuro-symbolic approaches [10]. Our results show that diverse APIs, granularity shifts, and multi-statement semantics make standalone strategies insufficient, motivating hybrid frameworks that combine deterministic heuristics with knowledge-driven or agentic reasoning. Our dataset and triple-isolation evaluation benchmark the resulting compositional gap [42].

## 8 Related Work

### 8.1 Template-Based API Replacement and Architecture-Level Refactorings

The most closely related work for detecting replacement opportunities is RETIWA (Replacing Custom Implementations With APIs) by Tufano et al. [50], which uses a curated knowledge base of code templates to identify handwritten logic matching predefined patterns. Although effective for well-structured, template-like cases, RETIWA relies on static matching and does not generalize to arbitrary expressions or composite, multi-statement logic. Another relevant approach is RefacLib by Taneja et al. [48], which automatically detects refactorings in evolving library APIs to support client migration. While effective at identifying structural API changes within libraries, RefacLib does not address the complementary problem of recommending client-side replacements of bespoke logic with existing APIs. Beyond code-level replacement, Stocker and Zimmermann [47] introduced the Interface Refactoring Catalog (IRC), which systematizes 22 architectural API refactorings addressing endpoint structure and evolution. These approaches, however, focus on either template-like cases, library evolution, or architectural API design rather than client-side, multi-statement logic replacement. To address this gap, our work adopts a hybrid, multi-phase methodology. We first mine and manually validate API replacement refactorings to construct a ground-truth dataset. We then synthesize structural and semantic features to derive deterministic heuristics for recurring, well-structured patterns. Finally, we introduce an adaptive knowledge discovery and retrieval framework that performs semantic reasoning over composite, multi-statement, and context-dependent code, enabling context-aware recommendations even when no predefined template exists.

### 8.2 API Recommendation and Library Selection Techniques

Traditional API recommendations focus on real-time synthesis during active development. Heinemann and Hummel [21] introduced an identifier-based technique that relies on semantic cues embedded in variable and method names, enabling recommendations even when APIs are absent in surrounding code. This distinguishes their work from earlier, structurally driven models that depended on type hierarchies or usage patterns. Nguyen et al. [33] advanced this line with APIREC, a statistical learning approach that mines fine-grained edits to identify repetitive change patterns. At the library selection level, CrossRec [34] assists in selecting third-party libraries based on dependency similarities, outperforming earlier tools like LibFinder [39] and LibCUP [45]. Across these studies, the underlying assumption is that developers seek guidance when choosing an API, however, they do not address the post-development refactoring scenario, which involves identifying where handwritten logic has already been implemented and can be replaced by more efficient API calls through refactoring.

### 8.3 Code Completion, Usage Ranking, and Argument Prediction

Building on type-aware API suggestions in modern IDEs, Bruch et al. [8] and Hou and Pletcher [22] proposed learning-based ranking models to prioritize relevant API calls. Huang et al. [23] incorporated external knowledge from Stack Overflow to further improve completion suggestions, while Chen et al. [12] developed neural architectures that integrate token and data-flow information to model correlated API-usage behaviors. Nguyen et al. [35] introduced ARist, which predicts API arguments by combining program analysis with learned representations of parameter semantics. While these techniques excel at assisting developers in completing API calls, they are not designed to detect semantically equivalent handwritten logic in existing code. Consequently, they provide ranking and synthesis support but lack the detection capabilities required for API replacement refactoring. Our work addresses this gap by conducting an empirical study to reveal the feature characteristics of such custom implementations and employing AKIRA, a hybrid adaptive knowledge discovery and Retrieval framework to detect API replacement opportunities.

### 8.4 AI-Assisted and LLM-Driven Refactoring Support

Recent advances in deep learning have enabled intelligent support for software engineering tasks such as code generation, defect prediction, program repair, API recommendation, and automated refactoring. Chen et al. [13] noted that, despite this progress, many approaches still struggle with semantic understanding, contextual reasoning, explainability, and robust deployment in real-world environments. These challenges are particularly relevant to API replacement refactoring, which requires reasoning about the semantic equivalence between handwritten multi-statement implementations and reusable library abstractions. Within the refactoring domain, Aniche et al. [4] explored supervised models for predicting refactorings, while Alomar et al. [3] and Chavan et al. [11] examined the impact of ChatGPT on developer practices. Pomian et al. [40] integrated LLMs into IDEs for Extract Method recommendations, though with limited explainability, and Liu et al. [29] showed that general-purpose LLMs struggle to reliably detect refactoring opportunities without substantial guidance. Li et al. [28] further demonstrated the importance of deeper semantic modeling for API recommendation. Building on these insights, AKIRA combines deterministic heuristics with a repository-grounded adaptive retrieval framework to perform context-sensitive reasoning over multi-statement code blocks, enabling the detection of both syntactic and semantic API replacement refactorings.

## 9 Conclusion

This paper presents the first systematic empirical study of API replacement refactorings, identifying them as a recurring yet underexplored practice in open-source Java applications. Through manual validation of 366 instances across six diverse projects, we demonstrated that these refactorings improve code maintainability but are difficult to detect with traditional heuristic patterns. To address this, we proposed AKIRA, a hybrid adaptive knowledge discovery and retrieval framework. By combining deterministic heuristics with a refactoring-aware knowledge base, AKIRA effectively discovers semantic API equivalents of ad hoc logic, outperforming current baselines in both precision and recall. Future work includes extending AKIRA to handle composite, inter-procedural replacements, integrating formal verification to ensure behavior preservation, and expanding the knowledge base to support cross-language API migration, enabling idiomatic library translation across software ecosystems.

## 10 Data Availability

Our replication package, including the dataset and supporting code, is available at [37].

## Acknowledgments

We thank the anonymous reviewers for their constructive feedback. This work was partially supported by the National Natural Science Foundation of China (62502343, 62232003, and 62572046).

## References

- [1] Miltiadis Allamanis and Charles Sutton. 2013. Mining source code repositories at massive scale using language modeling. In *2013 10th Working Conference on Mining Software Repositories (MSR)*. 207–216. doi:10.1109/MSR.2013.6624029
- [2] Eman Alomar, Mohamed Wiem Mkaouer, and Ali Ouni. 2019. Can Refactoring Be Self-Affirmed? An Exploratory Study on How Developers Document Their Refactoring Activities in Commit Messages. In *2019 IEEE/ACM 3rd International Workshop on Refactoring (IWoR)*. 51–58. doi:10.1109/IWoR.2019.00017
- [3] Eman Abdullah Alomar, Anushkrishna Venkatakrishnan, Mohamed Wiem Mkaouer, Christian D. Newman, and Ali Ouni. 2024. How to Refactor this Code? An Exploratory Study on Developer-ChatGPT Refactoring Conversations. In *2024 IEEE/ACM 21st International Conference on Mining Software Repositories (MSR)*. 202–206.
- [4] Maurício Aniche, Erick Maziero, Rafael Durelli, and Vinicius H. S. Durelli. 2022. The Effectiveness of Supervised Machine Learning Algorithms in Predicting Software Refactoring. *IEEE Transactions on Software Engineering* 48, 4 (2022), 1432–1450. doi:10.1109/TSE.2020.3021736
- [5] Apache Software Foundation. 2024. Apache Ant. https://github.com/apache/ant.
- [6] Apache Software Foundation. 2025. Apache Struts Web Application Framework. https://github.com/apache/struts
- [7] Jet Brains. 2025. *Intellij IDEA Community Edition*. https://www.jetbrains.com/idea/
- [8] Marcel Bruch, Martin Monperrus, and Mira Mezini. 2009. Learning from Examples to Improve Code Completion Systems. In *Proceedings of the 7th Joint Meeting of the European Software Engineering Conference and the ACM SIGSOFT Symposium on the Foundations of Software Engineering (ESEC/FSE '09)*. ACM, 213–222. doi:10.1145/1595696.1595728
- [9] Max Brunsfeld. 2018. Tree-sitter: An incremental parsing system for programming tools. https://github.com/tree-sitter/tree-sitter
- [10] Swarat Chaudhuri, Kevin Ellis, Oleksandr Polozov, Rishabh Singh, Armando Solar-Lezama, and Yisong Yue. 2021. Neurosymbolic Programming. *Foundations and Trends® in Programming Languages* 7, 3 (2021), 158–243. doi:10.1561/2500000049
- [11] Omkar Sandip Chavan, Divya Hinge, Soham Deo, Yaxuan Olivia Wang, and Mohamed Wiem Mkaouer. 2024. Analyzing Developer-ChatGPT Conversations for Software Refactoring: An Exploratory Study. *2024 IEEE/ACM 21st International Conference on Mining Software Repositories (MSR)* (2024), 207–211. https://doi.org/10.1145/3643991.3645082
- [12] Chi Chen, Xin Peng, Zhenchang Xing, Jun Sun, Xin Wang, Yifan Zhao, and Wenyun Zhao. 2022. Holistic Combination of Structural and Textual Code Information for Context Based API Recommendation. *IEEE Transactions on Software Engineering* 48, 8 (2022), 2987–3009. doi:10.1109/TSE.2021.3074309
- [13] Xiangping Chen, Xing Hu, Yuan Huang, He Jiang, Weixing Ji, Yanjie Jiang, Yanyan Jiang, Bo Liu, Hui Liu, Xiaochen Li, Xiaoli Lian, Guozhu Meng, Xin Peng, Hailong Sun, Lin Shi, Bo Wang, Chong Wang, Jiayi Wang, Tiantian Wang, Jifeng Xuan, Xin Xia, Yibiao Yang, Yixin Yang, Li Zhang, Yuming Zhou, and Lu Zhang. 2025. Deep Learning-Based Software Engineering: Progress, Challenges, and Opportunities. *SCIENCE CHINA Information Sciences* 68, 1 (2025), 111102. doi:10.1007/s11432-023-4127-5
- [14] Tangled Commit. 2024. *Guava*. https://github.com/google/guava/commit/f617e81434173fdedd4710b5c2f010d074379104
- [15] Danny Dig and Ralph Johnson. 2006. How do APIs evolve? A story of refactoring. *Journal of Software Maintenance and Evolution: Research and Practice* 18, 2 (March 2006), 83–107. doi:10.1002/smr.328
- [16] Elasticsearch. 2025. Elasticsearch. https://github.com/elastic/elasticsearch
- [17] The Eclipse Foundation. 2025. *EClipse IDE*. https://eclipseide.org/
- [18] Martin Fowler. 2018. *Refactoring: Improving the Design of Existing Code* (2nd ed.). Addison-Wesley Professional, Boston, MA. 448 pages.
- [19] Google LLC. 2025. Guava: Google Core Libraries for Java. https://github.com/google/guava
- [20] Aric A. Hagberg, Daniel A. Schult, and Pieter J. Swart. 2008. Exploring Network Structure, Dynamics, and Function using NetworkX. In *Proceedings of the 7th Python in Science Conference (SciPy2008)*. Pasadena, CA, USA, 11–15. https://doi.org/10.25080/TCWV9851
- [21] Lars Heinemann and Benjamin Hummel. 2011. Recommending API Methods Based on Identifier Contexts. In *Proceedings of the 3rd International Workshop on Search-Driven Development: Users, Infrastructure, Tools and Evaluation (SUITE '11)*. ACM, Honolulu, HI, USA. doi:10.1145/1985429.1985430
- [22] Daqing Hou and David M. Pletcher. 2010. Towards a Better Code Completion System by API Grouping, Filtering, and Popularity-Based Ranking. In *Proceedings of the 2nd International Workshop on Recommendation Systems for Software Engineering (RSSE '10)*. ACM, 26–30. doi:10.1145/1808920.1808926
- [23] Qiao Huang, Xin Xia, Zhenchang Xing, David Lo, and Xinyu Wang. 2018. API Method Recommendation without Worrying about the Task-API Knowledge Gap. In *Proceedings of the 33rd IEEE/ACM International Conference on Automated Software Engineering (ASE '18)*. IEEE / ACM, 293–304. doi:10.1145/3238147.3238191
- [24] Jeff Johnson, Matthijs Douze, and Hervé Jégou. 2021. Billion-Scale Similarity Search with GPUs. *IEEE Transactions on Big Data* 7, 3 (2021), 535–547. doi:10.1109/TBDATA.2019.2921572
- [25] JUnit5. 2025. JUnit5. https://github.com/junit-team/junit5

[26] Eirini Kalliamvakou, Georgios Gousios, Kelly Blincoe, Leif Singer, Daniel M. German, and Daniela Damian. 2014. The Promises and Perils of Mining GitHub. In *Proceedings of the 11th Working Conference on Mining Software Repositories (MSR 2014)*. ACM, 92–101. doi:10.1145/2597073.2597074

[27] Commons Lang. 2025. Commons Lang. https://github.com/apache/commons-lang

[28] Zhihao Li, Chuanyi Li, Ze Tang, Wanhong Huang, Jidong Ge, Bin Luo, Vincent Ng, Ting Wang, Yucheng Hu, and Xiaopeng Zhang. 2024. PTM-APIRec: Leveraging Pre-trained Models of Source Code in API Recommendation. *ACM Transactions on Software Engineering and Methodology* 33, 3 (2024), 72:1–72:30. doi:10.1145/3632745

[29] Bo Liu, Yanjie Jiang, Yuxia Zhang, Nan Niu, Guangjie Li, and Hui Liu. 2025. Exploring the potential of general purpose LLMs in automated software refactoring: an empirical study. *Automated Software Engineering* 32, 1 (2025), 26. doi:10.1007/s10515-025-00500-0

[30] Microsoft Corporation. 2025. Visual Studio Code. https://code.visualstudio.com.

[31] Emerson Murphy-Hill, Chris Parnin, and Andrew P. Black. 2012. How We Refactor, and How We Know It. *IEEE Transactions on Software Engineering* 38, 1 (2012), 5–18. doi:10.1109/TSE.2011.41

[32] Munkhtsetseg Namsraidorj, Sanchirjav Lkhaasuren, Bagabold Gendensuren, Khantulga Radnaa, Javkhlan Rentsendorj, and Amirlan Enkhtur. 2024. Continuous Integration and Delivery of Software products: Pipeline implementation. *International Journal of Engineering and Computer Science* (2024). https://ijecs.in/index.php/ijecs/article/view/4821

[33] Anh Tuan Nguyen, Michael Hilton, Mihai Codoban, Hoan Anh Nguyen, Lily Mast, Eli Rademacher, Tien N. Nguyen, and Danny Dig. 2016. API Code Recommendation using Statistical Learning from Fine-Grained Changes. In *Proceedings of the 24th ACM SIGSOFT International Symposium on Foundations of Software Engineering (FSE 2016)*. 511–522. doi:10.1145/2950290.2950333

[34] Phuong T. Nguyen, Juri Di Rocco, Davide Di Ruscio, and Massimiliano Di Penta. 2020. CrossRec: Supporting software developers by recommending third-party libraries. *Journal of Systems and Software* 161 (2020), 110460. doi:10.1016/j.jss.2019.110460

[35] Son Nguyen, Cuong Tran Manh, Kien T. Tran, Tan M. Nguyen, Thu-Trang Nguyen, Kien-Tuan Ngo, and Hieu Dinh Vo. 2023. ARist: An effective API argument recommendation approach. *Journal of Systems and Software* 204 (2023), 111786. doi:10.1016/j.jss.2023.111786

[36] Bridget Nyirongo, Yanjie Jiang, Nan Niu, and Hui Liu. 2025. An Empirical Study of Software Refactorings in Real-World Open-Source Java Projects. *IEEE Transactions on Software Engineering* 51, 11 (2025), 3013–3037. doi:10.1109/TSE.2025.3604821

[37] Bridget Nyirongo, Yanjie Jiang, Yuxia Zhang, and Hui Liu. 2026. *Replication Package: From Custom Logic to APIs: Understanding and Recommending API Replacement Refactorings*. doi:10.5281/zenodo.20095466

[38] OpenAI. 2024. GPT-4o System Card. https://openai.com/index/gpt-4o-system-card/

[39] Ali Ouni, Raula Gaikovina Kula, Marouane Kessentini, Takashi Ishio, Daniel M. German, and Katsuro Inoue. 2017. Search-based software library recommendation using multi-objective optimization. *Information and Software Technology* 83 (2017), 55–75. doi:10.1016/j.infsof.2016.11.007

[40] Dorin Pomian, Abhiram Bellur, Malinda Dilhara, Zarina Kurbatova, Egor Bogomolov, Timofey Bryksin, and Danny Dig. 2024. Next-Generation Refactoring: Combining LLM Insights and IDE Capabilities for Extract Method. In *2024 IEEE International Conference on Software Maintenance and Evolution (ICSME)*. 275–287. doi:10.1109/ICSME58944.2024.00034

[41] False Positive. 2024. *Springframework*. https://github.com/spring-projects/spring-framework/commit/80b264ba8201256d55350eeeb4cdb40f73528cc6

[42] Ofir Press, Muru Zhang, Sewon Min, Ludwig Schmidt, Noah A. Smith, and Mike Lewis. 2022. Measuring and Narrowing the Compositionality Gap in Language Models. *ArXiv* abs/2210.03350 (2022). https://api.semanticscholar.org/CorpusID:252762102

[43] QuarkusIO community. 2025. Quarkus: Supersonic Subatomic Java Framework. https://github.com/quarkusio/quarkus

[44] Soumaya Rebai, Marouane Kessentini, Vahid Alizadeh, Oussama Ben Sghaier, and Rick Kazman. 2020. Recommending refactorings via commit message analysis. *Inf. Softw. Technol.* 126 (2020), 106332. https://doi.org/10.1016/j.infsof.2020.106332

[45] Mohamed Aymen Saied, Ali Ouni, Houari Sahraoui, Raula Gaikovina Kula, Katsuro Inoue, and David Lo. 2018. Improving reusability of software libraries through usage pattern mining. *Journal of Systems and Software* 145 (2018), 164–179. doi:10.1016/j.jss.2018.08.032

[46] Springframework. 2025. Springframework. https://github.com/spring-projects/spring-framework

[47] Mirko Stocker and Olaf Zimmermann. 2023. API Refactoring to Patterns: Catalog, Template and Tools for Remote Interface Evolution. In *Proceedings of the 28th European Conference on Pattern Languages of Programs (EuroPLoP 2023)*. ACM, 2:1–2:32. doi:10.1145/3628034.3628073

[48] Kunal Taneja, Danny Dig, and Tao Xie. 2007. Automated Detection of API Refactorings in Libraries. In *Proceedings of the 22nd IEEE/ACM International Conference on Automated Software Engineering (ASE'07)* (Atlanta, Georgia, USA). IEEE/ACM, 377–380. doi:10.1145/1321631.1321688

[49] Hugo Touvron, Thibaut Lavril, Gautier Izacard, Xavier Martinet, Marie-Anne Lachaux, Timothée Lacroix, Baptiste Rozière, Naman Goyal, Eric Hambro, Faisal Azhar, Aur'elien Rodriguez, Armand Joulin, Edouard Grave, and Guillaume Lample. 2023. LLaMA: Open and Efficient Foundation Language Models. *ArXiv* abs/2302.13971 (2023). https://api.semanticscholar.org/CorpusID:257219404

[50] Rosalia Tufano, Emad Aghajani, and Gabriele Bavota. 2022. Don't Reinvent the Wheel: Towards Automatic Replacement of Custom Implementations with APIs. In *2022 IEEE International Conference on Software Maintenance and Evolution (ICSME)*. 394–398. doi:10.1109/ICSME55016.2022.00046